\documentclass[]{rsos}

\begin{document}

\title{Extragalactic Background Light Measurements and Applications}

\author{
Asantha Cooray}

\address{Department of Physics \& Astronomy, University of California, Irvine, CA 92697}

\subject{Astronomy}


\corres{Insert corresponding author name\\
\email{acooray@uci.edu}}

\begin{abstract}
This review covers the measurements related to the extragalactic background light (EBL) intensity
from gamma-rays to radio in the electromagnetic spectrum over 20 decades in the wavelength. 
The cosmic microwave background (CMB) remains the best measured spectrum with an accuracy better than 1\%. 
The measurements related to the cosmic optical background (COB), centered
at 1 $\mu$m, are impacted by the large zodiacal light  associated with interplanetary dust in the inner
Solar system. The best measurements of COB come from an indirect technique involving 
Gamma-ray spectra of bright blazars with an absorption feature resulting from pair-production off of COB photons.
The cosmic infrared background (CIB) peaking at around 100 $\mu$m established
an energetically important background with an intensity comparable to the optical background. This discovery
paved the path for large aperture far-infrared and sub-millimeter observations resulting in the
discovery of dusty, starbursting galaxies. Their role in galaxy formation and evolution remains an active area of research
in modern-day astrophysics. The extreme UV background remains mostly unexplored and will be a challenge to measure
due to the high Galactic background and absorption of extragalactic photons by the intergalactic medium 
at these EUV/soft X-ray energies. We also summarize our understanding of the spatial anisotropies and angular power spectra of intensity fluctuations.
We motivate a precise direct measurement of the COB between 0.1 to 5 $\mu$m  using a small aperture telescope
observing either from the outer Solar system, at distances of 5 AU or more, or out of the ecliptic plane. Other future applications
include improving our understanding of the background at TeV energies and spectral distortions of CMB and CIB.
\end{abstract}

\maketitle

\section{Introduction}

The extragalactic background light (EBL) is the integrated intensity of all of the
light emitted throughout the history of the universe across the whole of the electromagnetic
spectrum. While EBL is sometimes defined as the extragalactic intensity spectrum from UV to infrared (e.g., see
review in Dwek \& Krennrich 2012), the total energy content of the universe in the electromagnetic spectrum spans close to 20 decades
in the wavelength from gamma-rays to radio. Across this whole range, the EBL spectrum
captures cosmological backgrounds associated with either primordial phenomena, such as the
cosmic microwave background (CMB), or photons emitted by stars, galaxies, and active galactic
nuclei (AGN) due to nucleosynthesis or other radiative processes, including dust scattering, absorption and reradiation. 
The EBL may also contain signals that are diffuse and extended, including high energy photons associated with
dark matter particle decays or annihilation.

In the UV to infrared portion of the electromagnetic spectrum, the EBL spectrum captures
the redshifted energy released from all stars and galaxies throughout the cosmic history, including
first stellar objects, primordial black holes, and  proto-galaxies.
If precisely measured, the EBL spectrum can be used as a constraint on models of galaxy formation and evolution, 
while providing an anchor that connects global radiation energy density to star formation, metal production, and 
gas consumption. The microwave background spectrum at microns to mm-wave radio wavelengths, associated with CMB 
photons, has been measured to a precision better than a percent and is described by
a black-body spectrum with a temperature of $2.7260 \pm 0.0013$ (Fixsen 2009). 
Such a measurement is facilitated by  the fact that the CMB is the brightest of the EBL components with a
factor of 30 to 40 higher energy density than the next brightest background at 
optical to infrared wavelengths.  The CMB is also a well-known probe of cosmology.
The anisotropies come from both primordial physics, at the epoch of last scattering when electrons and protons first
combined to form hydrogen, or secondary effects during the propagation of photons. The latter includes
effects associated with both gravitational and scattering effects. The angular power spectrum of CMB spatial anisotropies 
has now been measured down to a few arcminute scales with Planck and has been used to determine cosmological parameters such as the
energy density contents, the spatial curvature, spectral index of primordial density perturbations laid
out after an inflation epoch, among others.

Despite the limitations on the accuracy of existing EBL intensity measurements 
there have been some key breakthroughs due to intensity measurements of 
the sky. A classic example is the cosmic IR background (CIB) peaking at 100 $\mu$m. Its intensity was measured with
instruments such as DIRBE (Hauser et al. 1998) and at wavelengths beginning 250 $\mu$m with FIRAS (Fixsen et al. 1998), 
both on COBE. The EBL intensity peaking at 100 microns was 
found to be roughly comparable to that of the optical background, suggesting that CIB at long wavelengths is 
energetically important as the optical/near-IR background dominated by galaxies. This then motivated
high resolution far-infrared and sub-millimeter imaging from both space and ground and with increasing
aperture sizes and sensitivity more of the CIB background has been resolved to point sources. These point
sources are mainly dusty, star forming and starbursting galaxies at high redshifts (see review in Casey et al. 2014). 
Their role in galaxy formation and galaxy evolution remains one of the key topics in sub-mm astronomical observations today. 

With a precise measurement of the EBL intensity spectrum,
a cosmic consistency test can be performed  as a function of the wavelength
by comparing the integrated light from all galaxies, stars, AGN and other point sources,
to the EBL intensity. Any discrepancies suggest the presence of new, diffuse emission that is
unresolved by telescopes. The possibilities for new discoveries with profound implications for astronomy range from
 recombination signatures during reionization and diffuse photons
associated with dark matter annihilation and their products. Related to the last scientific possibility important
studies have been carried out, with more expected over the coming years, whether there is a diffuse signature 
at GeV energy scales in the cosmic gamma-ray background (CGB) as measured by the Fermi-LAT that can be ascribed to 
dark matter (e.g., Ando \& Ishiwata 2015).

In addition to the total EBL intensity significant information on the sources of emission and their nature comes from
measurements that focus on the anisotropies of the intensity across the sky. These are
in general quantified and measured in terms of the angular power spectrum or the correlation function.
The well-studied example in the literature is the anisotropy power spectrum of the CMB, resulting in
high-precision cosmological parameter measurements (e.g., Ade et al. 2015a). Anisotropy power spectra have also been measured for
CIB, COB, and CGB leading to inferences on the properties of the source populations present at these
wavelengths, especially on certain physical details related to the
faint sources that are below the individual point source detection level.

The existing EBL intensity measurements are due to a combination of ground and space-based observations of the sky.
Direct absolute intensity measurements  must account  for a variety of
foregrounds both within the Solar system, such as Zodiacal light at optical and infrared wavelengths, to
Milky Way, such as the Galactic emission at radio, infrared, X-ray or gamma rays. A good example of an indirect technique 
to measure EBL is the use of absorbed TeV spectra of individual blazars and other AGNs at cosmological distances to infer
the number density of intervening infrared photons that are responsible of
electron-positron pair-production by interactions with TeV photons. This has led to the best determined
COB measurements in the literature, especially given the fact that modeling and removing Zodiacal
light remains a challenge for direct EBL intensity measurements around 1 $\mu$m.

\begin{figure}[!t]
\begin{center}
\includegraphics[width=0.9\textwidth]{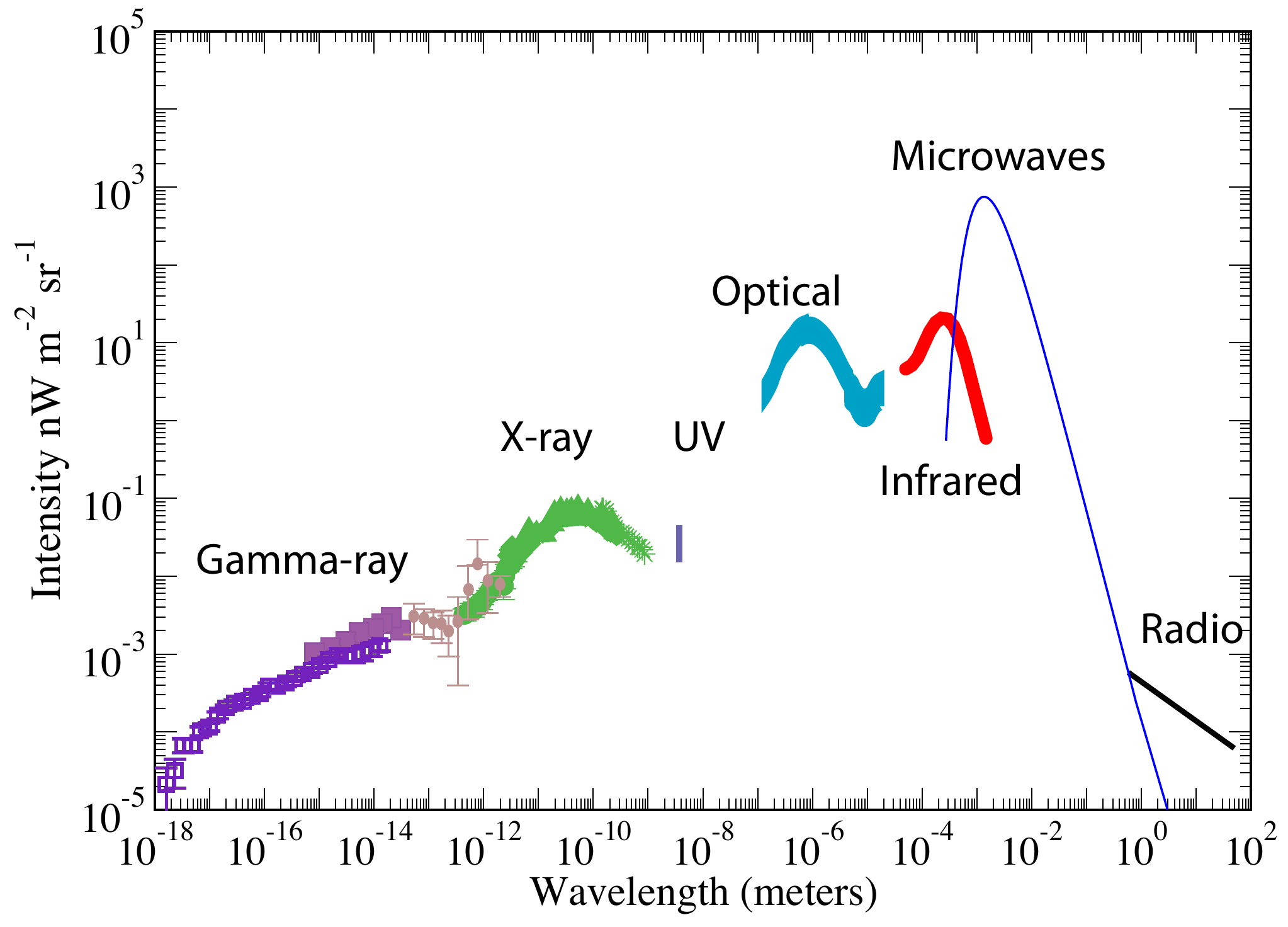}
\end{center}
\caption{
Intensity of the extragalactic background ($\nu I_\nu$ in units of nW m$^{-2}$ sr$^{-1}$) as a function of the wavelength in meters.
We combine the existing measurements from the literature to highlight the  best determined estimates for the background
from gamma-ray to radio. The cosmic microwave background (CMB) has the least uncertainty as the spectrum is determined
to better than 1\%. Cosmic optical background (COB) has large uncertainties involving direct measurements due to uncertain
removal of the Zodiacal light foreground. Here we show the indirect estimate of EBL at optical wavelengths based on
the TeV/gamma-ray absorption spectra of distant blazars. The UV/soft X-ray background at a wavelength of 10 to 100 nm
remains unexplored. From left to right in increasing wavelength, the plotted datasets are: Fermi-LAT (the total extragalactic background composed
of diffuse and resolved point sources; Ackermann et al. 2015) and EGRET (Sreekumar et al. 1998; we have removed three data points from Strong et al. 2004
at highest energies) in the Gamma-ray spectrum, COMPTEL (filled circles between Gamma-ray and X-rays; Weidenspointner et al. 2000)
between Gamma-ray and X-rays, HEAO1 A2 and A4 (Marshall et al. 1980; Rothschild et al. 1983), 
INTEGRAL (Churazov et al. 2007), SWIFT/BAT (Ajello et al. 2008), Nagoya balloon experiment (Fukada et al. 1975),
SMM (Watanabe et al. 2000), ASCA (Gendreau et al. 1995) and RXTE (Revnivtsev et al. 2003) in the hard to soft X-ray regime in green
symbols, DXS and CHIPS in soft X-rays/Extreme UV (as discussed in Smith et al. 2014 as a line at 0.25 KeV), 
HESS in optical (Abramowski et al. 2012; see Figure 2 for other measurements), DIRBE (Hauser et al. 1998) and FIRAS (Lagache et al. 2000)
in the far-infrared, FIRAS at microwaves (Mather et al. 1994; Fixsen et al. 1996) and ARCADE (Kogut et al. 2011) in the radio.
The area under each of these backgrounds capture the total energy density of the photons in each of those
wavelength regimes. From gamma-rays to radio the integrated intensity values 
in units of nW m$^{-2}$ sr$^{-1}$ for key EBL components are $\sim 0.015$ (gamma-ray), $\sim 0.3$ (X-ray), 
$0.01-0.02$ (lower and upper limits at 4.9 nm for Extreme UV), $24 \pm 4$ (with an additional $\pm 5$ systematic; optical),
$\sim 30 \pm 10$ (CIB), $960$ (CMB), and $<0.001$ (radio), respectively.
}
\end{figure}

We summarize existing EBL intensity measurements in Figure~1 where we plot the spectral intensity $\lambda I_\lambda$
as a function of the wavelength $\lambda$. In this figure the area under each of the
spectral components represents the total energy density associated with each of those backgrounds. Those values are listed in the caption of
Figure~1 where the estimates were made using a statistical average of existing results from the literature.
In most of these measurements large systematics, associated with foreground models, are likely to be still present. 
 Here we briefly outline the techniques, foregrounds, and systematics associated with EBL measurements. We also discuss their applications for
astrophysical and cosmological studies and briefly summarize studies related to spatial anisotropies.
We cover from short to long wavelengths starting from the gamma-ray background.

\subsection{Gamma-Ray}

The early measurements of the Cosmic Gamma-Ray Background (CGB) intensity came from SAS-2 between 40 and 300 MeV in 1978 (Fichtel et al.
1978), followed by EGRET between 40 Mev and 10 Gev in 1998 (Sreekumar et al. 1998; Strong et al. 2004). These measurements have 
been superseded in this decade by Fermi-LAT covering 100 MeV to 800 GeV with roughly 25-30 times better sensitivity than EGRET,
as well as overall an improvement in the flux calibration. The CGB spectrum measured by Fermi-LAT shows a cutoff at energy scales around 280 GeV (Ackermann et al. 2015). 
Below this cut-off  the spectrum  can be described by a single power-law with a spectral index about 2.3 ($\pm 0.05$). The cutoff is explained as the disappearance
of the high energy photons that are pair-producing via interactions with the infrared background photons that
we discuss later (Aharonian et al. 2006; Ackermann et al. 2012a; Dominguez et al. 2011; Gilmore et al. 2012).

The CGB spectrum below the cutoff is mostly explained in terms of a combination of AGNs in the form of blazars and gamma-ray
emission from star-forming galaxies (SFGs). Small, but non-negligible depending on the exact energy, comes
from millisecond pulsars (MSPs), Type Ia supernovae, and gamma-rays from galaxy clusters. 
At energies above $\sim$ 50 Gev blazars are able to fully account for the background, with an  estimate of $86^{+16}_{-14}$\% in
explained by extrapolated blazar counts in Fermi-LAT Collaboration (2015).
Between 0.1 and 50 Gev, blazars account for about 20\% of the CGB. The rest are due to
other populations of AGNs and SFGs. We refer the reader to a comprehensive review by
Fornasa and S\'anchez-Conde (2015) on the source populations contributing to
CGB and existing population models in the literature.

The literature also considers the possibility of a dark matter-induced signal in the gamma-ray background (see Bertone et al. 2005 for a review).
This could be from dark matter that decays into standard particles (e.g., Bertone et al. 2007; Ibarra et al. 2008)
or due to annihilation products (Abdo et al. 2010; see also 
Palomares-Ruiz \& Siegal-Gaskins 2010). These gamma-rays will form an anisotropic signal associated with dark matter in our Galaxy.
However, the signal is also expected to be present in all galaxies. The integrated signal from  dark matter halos of all other galaxies
will lead to a clustering signal in the gamma-ray background and must be separated from the clustering of all gamma-ray emitting faint extraglactic 
sources that also contribute to the background.
A direct detection of a signal associated with dark matter decay has been attempted towards Galactic center (e.g., Abazajian \& Kaplinghat 2012) 
and nearby dwarf galaxies that are considered to be dark matter rich (e.g., Drlica-Wagner et al. 2015, Geringer-Sameth et al. 2015). The claims of 
gamma-ray excesses toward the Galactic center
have been questioned as whether due to pulsars or other astrophysical foregrounds (e.g., Bartels et al. 2015;  O'Leary et al. 2015),
while the signal towards dwarf galaxies remain at the level of a 3$\sigma$ detection (e.g., Hooper \& Linden 2015).

Due to spatial resolution and the all-sky nature of the CBG measurements, Fermi-LAT  also provides a treasure
trove of data beyond the energy spectrum. In particular anisotropies or spatial fluctuations of the CGB have been
pursued to study the nature of faint sources that can account for the small diffuse signals in the CBG below the point
source detection of current high energy instruments. 
The angular power spectrum of the CGB is mostly Poisson or shot-noise like between multipole ell ranges of 150 to 500, corresponding to
30 arcminutes to 3 degree angular scales on the sky (e.g., Ackermann et al. 2012b). 
Moving beyond auto spectra, cross-correlations of the anisotropies can also be pursued (e.g., Xia et al. 2011).
For an example the diffuse CGB signal from all of decaying dark matter
in the universe can be studied via cross-correlations with large-scale structure tracers of the same dark matter,
such as weak lensing maps (e.g., Camera et al. 2013).  Existing cross-correlation attempts using Fermi-LAT data include
both the CMB lensing map from Planck (Fornengo et al. 2015) and galaxy weak lensing in
the CFHT Lensing Survey (Shirasaki et al. 2014). These studies are likely to be improved in the near future with
wider area galaxy lensing surveys such as those expected from the Dark Energy Survey (DES) and Large Synoptic Survey Telescope (LSST).

Over the next decade, CGB studies will be extended to higher energies with the 
Cherenkov Telescope Array (CTA) with detections likely in 10 Gev to 10 TeV energy range (Acharya et al. 2013). 
CTA will allow studies related to CBG fluctuations at higher energies, especially TeV scales where there are still no
reliable measurements on the spectrum or intensity fluctuations.  We also lack a
a complete understanding of the sources that contribute to CGB at 1 to 10 MeV scales, below the 100 MeV sensitivity of
Fermi-LAT. This background spectrum based on EGRET and COMPTEL (Figure 1) suggests the possibility of a
smooth connection to the cosmic X-ray background (CXB) at 10 to 100 KeV energy scales, though subject to
large uncertainties in COMPTEL background light measurements. If continue to be confirmed such a smooth transition to X-rays
suggest a different source population for the MeV background than the background at GeV energies.
The leading possibility is a combination of Seyfert galaxies and flat-spectrum radio quasars that appear as bright 
MeV sources. Due to large uncertainties with EGRET and COMPTEL background measurements at these energy scales, however,
we recommend a future experiment to improve background intensity measurements at MeV energy scales.

\subsection{X-Ray}

Cosmic X-ray background (CXB) is generally divided into high and soft energies, with transition
around an energy scale of 2 keV. Early measurements of
the hard CXB intensity came from HEAO1 between 2-30 keV and 10-400 keV with A2 and A4 instruments.
These measurements showed that the spectrum follows:
$I(E) = 7.9\times10^{-0.29} \exp[-E/41.1 {\rm keV}\; {\rm keV cm^{-2} s^{-1} sr^{-1} keV^{-1}}$,
consistent with thermal bremsstrahlung radiation with a temperature $\sim$ 40 keV (e.g., Marshall et al. 1980; 
Rothschild et al. 1983; Garmire et al. 1992). 
Bulk of the energy density of CXB is thus at 30 keV, but understanding the sources at such a high X-ray
energy has been slow. Previous surveys with SWIFT/BAT and Integral (Ajello et al. 2008; Churazov et al. 2007)
only resolved 1\% of the background to point sources at 30 keV. Current measurements in this energy scale are mainly from
NuSTAR (Harrison et al. 2013).

Recent models show that in order to match both the redshift distribution of the faint X-ray sources in
deep images with the {\it Chandra} X-ray Observatory and the overall CXB spectrum requires an
evolving ratio of Type 1 (AGNs with visible nuclei) to Type 2 (with obscured broad-line regions) 
sources, such that there are more Type 2 Seyferts at higher redshifts 
(e.g., Menci et al. 2004; Ueda et al. 2003; Gilli et al. 2007). Deep surveys with NuSTAR
resolve 35\% of the 8-24 KeV background to AGNs with obscuring columns up to 10$^{25}$ cm$^{-2}$
(Harrison et al. 2015), consistent with expectations from AGN population synthesis models (Worsley et al. 2005). 
Additional evidence for the presence of such highly absorbed AGNs exist as IR-bright sources in AKARI with 50\% of the
mid-IR AGN samples currently undetected in deep Chandra surveys (e.g., Krumpe et al. 2014).

The soft CXB has been measured with ROSAT down to 0.1 keV energies. At such low energies
intensity measurements start to become challenging due to the Galactic signal and there is a clear indication
for thermal emission from hot gas with a temperature of 10$^6$K associated with a hot component of
the interstellar medium (ISM) or the Galactic halo. 
Deep ROSAT imaging data resolved 80\% of the CXB at soft X-ray energies of 1 keV to
to discrete sources, mainly  bright AGNs. Chandra X-ray Observatory, with
spatial resolution at the level of 0.5 arcsec, and XMM-Newton has resolved $>$
90\% of the X-ray background at energies between 0.5 to 2 keV 
and $>$ 80\% at hard 2 to 9 keV energies (Worsley et al. 2005; Xue et al. 2012).  
The dominant source is AGNs with a non-negligible contribution from galaxy clusters (e.g., Wu \& Xue 2001)
and starbursting galaxies (e.g., Persic \& Rephaeli 2003).
The uncertainty in the unresolved intensity
is not in the source population but in the overall normalization of the total XRB intensity. 

With the background resolved to individual sources, anisotropy measurements of the CXB intensity
and their applications have been limited when compared to similar studies at other wavelengths.
In principle anisotropy studies can uncover diffuse X-ray sources
or faint sources below the detection level. A recent example is the use of Chandra deep
images to study the X-ray background fluctuations in combination with fluctuations
measured with Spitzer at 3.6 $\mu$m (Cappelluti et al. 2013). This signal has been explained as due to
the infrared and X-ray emission from  direct collapse
black holes (DCBHs) during reionization (Yue et al. 2013). An X-ray surveyor, such as the planned {\it Athena} mission,
should facilitate more detailed studies on the diffuse background, faint unresolved sources in current deep surveys such
as DCBHs, and multi-wavelength cross-correlation studies. 

\subsection{UV}

EBL measurements at UV  wavelengths exist with GALEX at 150 nm (Murthy et al. 2010),
with Voyager 1 and 2 at 110 nm (Murthy et al. 1999), and with Voyager UVS at 100 nm (Edelstein et al. 2000),
though subject to both large statistical and systematic uncertainties. In the extreme UV (EUV) wavelengths below 100 nm, and down to 10 nm 
at energy scales of 0.1 KeV in soft X-rays, there are no useful measurements of the Cosmic UV background (CUVB) 
in the literature (Fig.~1).  While technological developments can be expected, a measurement of the extragalactic EUV
background will likely remain challenging due to absorption of the extragalactic
photons by neutral hydrogen in our Galaxy and the intergalactic medium at wavelengths below 91.2 nm. Furthermore
the Galactic soft X-ray/EUV background presents a considerable foreground that limits a reliable measurement of the
UV background. And the next best measurements are in X-rays from a wavelength of 5 nm corresponding to energies of 0.25 KeV.

While a reliable background intensity might be challenging, further EUV observations are warranted since it is understood that bulk
of the baryons exists in the warm intergalactic medium (Cen \& Ostriker 1999)
with signatures that involve emission and absorption
lines around 50 nm (Tepper-Garcia et al. 2013). Based on Extreme UltraViolet Explorer (EUVE) data, 
there are some indications that certain galaxy clusters like Coma 
show excess EUV emission, especially at energies between 65 and 200 eV (Lieu et al. 1996).  While the wavelength regime between
10 to 100 nm remains a crucial wavelength regime for further exploration, we recommend further attempts between 100 and 1000 nm
in the UV as there are possibilities for some significant discoveries on the nature of warm intergalactic medium.
Instruments on the planetary spacecraft to outer Solar system may continue to provide opportunities for UV background measurements,
similar to past attempts with Voyager (Murthy et al. 1999; Edelstein et al. 2000).
In this respect one near-term possibility would be the use of New Horizons' Alice UV spectrometer (Stern et al. 2007)
for a new background measurement  at wavelengths around 140-180 nm.

\begin{figure}[!t]
\begin{center}
\includegraphics[width=0.9\textwidth]{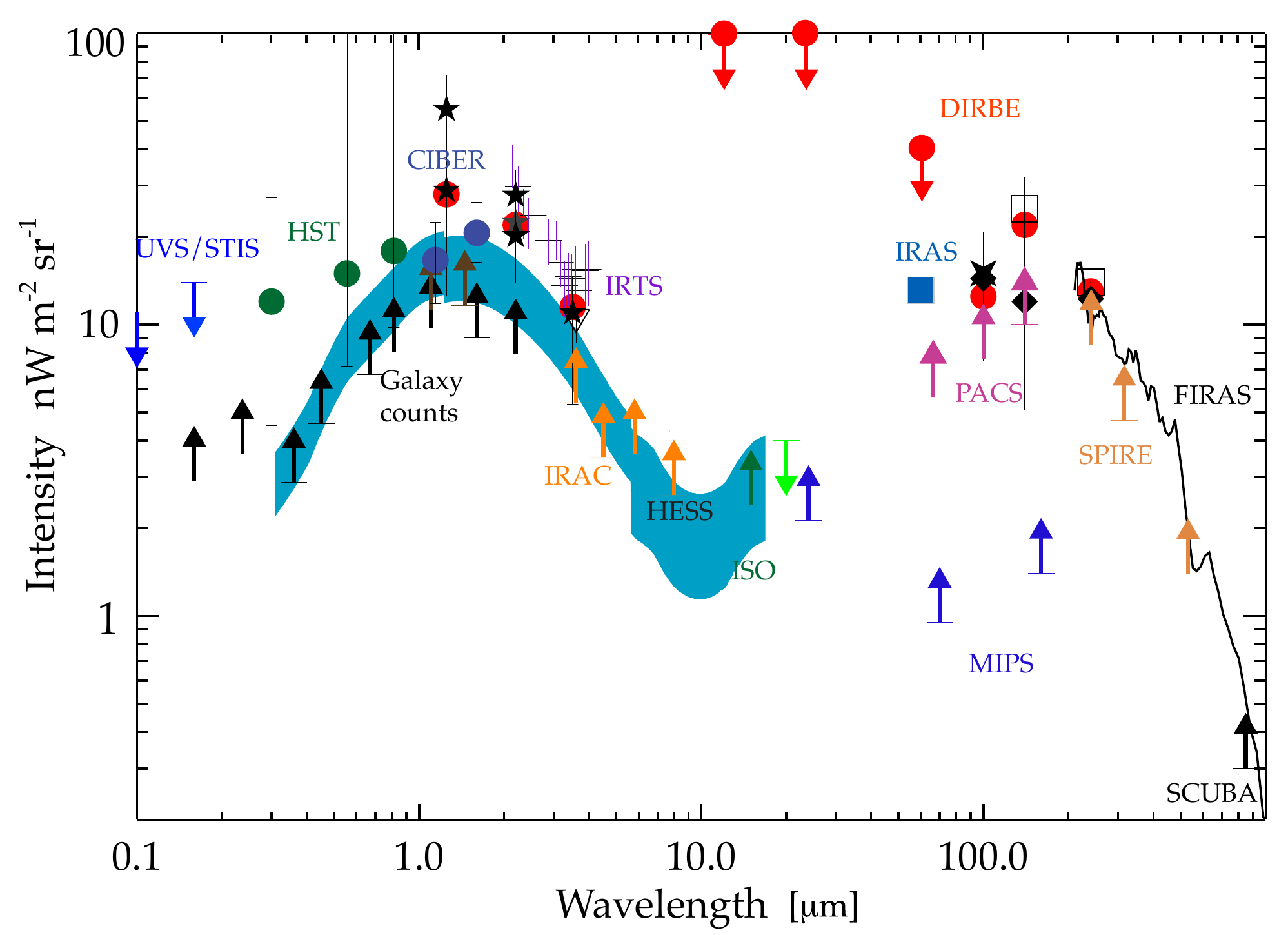}
\end{center}
\caption{
The cosmic optical and infrared background light from 0.1 to 100 $\mu$m.
The data points with error
bars are direct estimates using DIRBE (red circles: Wright 2004, Wright 2001; stars: Cambresy et al. 2001 at 1.25 and 2.2 $\mu$m; 
Gorjian et al. 2000 at 2.2 and 3.5 $\mu$m; Levenson et al. 2007 at 2.2 and 3.5 $\mu$m; open squares at 140 and 240 $\mu$m; Hauser et al. 1998), 
IRTS (purple crosses; Matsumoto et al. 2005), 
{\it Spitzer} at 3.6 $\mu$m (open triangle; Levenson \& Wright 2008), {\it Hubble} (green circles; Bernstein 2007), 
UVS/STIS (blue upper limits; Edelstein et al. 2000; Brown et al. 2000), CIBER (blue circles; model-dependent based on fluctuation
measurements; Zemcov et al. 2014), FIRAS (black line; with an overall uncertainty of 30\% between 200 $\mu$m and 1.2mm; Lagache et al. 2000
also Fixsen et al. 1998), and IRAS (blue square; 60 $\mu$m fluctuation-based estimate of EBL with IRAS;  Miville-Desch\^nes et al. 2002).
The lower limits to the EBL are from integrated or source counts using Hubble (Gardner et aL. 2000; Madau \& Pozzetti 2000), Spitzer/IRAC (Fazio et al. 2004), ISO (Elbaz et al. 1999), Spitzer/MIPS (Papovich et al. 2004; Dole et al. 2004),  Herschel/PACS (Berta et al. 2010), Herschel/SPIRE (B\'ethermin et al. 2012), and SCUBA (Smail et al. 2002).
The blue shaded region is the estimate of EBL using the HESS TeV blazar absorption spectra (Abramowski et al. 2012).
Apart from recent Herschel measurements, CIBER, and the estimate of EBL from HESS, all other measurements plotted here are tabulated in
Hauser \& Dwek (2011). This figure is based on a previous figure by Dole et al. (2006) that summarized these EBL and integrated galaxy count measurements.
}
\end{figure}

\subsection{Optical/Near-Infrared}

At optical and near-infrared (IR) wavelengths between 0.1 to 5 $\mu$m the EBL intensity is predominantly due to stellar
emission from nucleosynthesis throughout the cosmic history (see Hauser \& Dwek 2001 for a review).
The cosmic optical background (COB) spectrum also includes radiative information from the reionization epoch.
Due to redshifting of the UV photons to near-infrared emission from primordial sources is primarily at
wavelengths longward of 1 $\mu$m (Santos et al. 2002; Salvaterra et al. 2003). This includes
diffuse Ly-$\alpha$ and free-free radiation in addition to direct emission by stars and mini-quasars
(e.g., Cooray \& Yoshida 2004; Fernandez \& Komatsu 2006).

At optical and near-IR the few attempts at absolute measurements involve DIRBE on COBE in several band-passes between 1.25 $\mu$m and 
240 $\mu$m (Hauser et al. 1998; Cambresy et al. 2001; Gorjian et al. 2000; Levenson et al. 2007;  Wright 2001; Wright 2004),
IRTS, a small JAXA mission, between 1 and 4 $\mu$m (Matsumoto et al. 2005), Voyager (Edelstein et al. 2000) and
Hubble (Brown et al. 2000; Bernstein et al. 2002; Bernstein 2007).
Because DIRBEs confusion limit was 5th magnitude at 2.2 $\mu$m, 
all recent EBL measurements using DIRBE require subtraction of stellar light using ancillary measurements, such 
as 2MASS (Wright 2001; Levenson et al. 2007).  
While Hubble has been used for optical (Bernstein et al. 2002; Bernstein 2007) and far-UV (Brown et al. 2000) 
EBL measurements, the instrument was not designed for absolute photometry and required a 
careful subtraction of instrumental emission and baselines (e.g. dark current). 
Those measurements are subject to large uncertainties (e.g., Mattila 2003).

The dominant limitation for direct EBL intensity spectrum
at these wavelengths is the Zodiacal light associated with scattered Solar light from
micron-size dust interplanetary dust (IPD) particles near the Earth's orbit. Existing measurements with DIRBE on COBE
make use of model to remove Zodiacal light (Kelsall et al. 1998) or slight variations (Wright 2001).
The Zodiacal light foreground limits the accuracy of EBL intensity to roughly
an order of magnitude at wavelengths about 1 micron (Hauser et al. 1998; Gorjian et al. 2000;
Wright 2001; Cambresy et al. 2001; Matsumoto et al. 2005; Levenson \& Wright 2007).
Techniques to remove Zodiacal light includes monitoring of Fraunhofer lines in the dust scattered spectrum
relative to the Solar spectrum and use of the equivalent width of the lines to estimate the column
density of dust. In the case of
HST results on the optical background the zodiacal emission based on the observed strength of the reflected Fraunhofer lines from a ground-based 
measurement (Bernstein et al. 2002). The sounding rocket experiment CIBER (Zemcov et al. 2013) is capable of absolute photometry 
(Tsumura et al. 2013) and results related to the optical/near-IR EBL are soon expected (Matsuura et al. in preparation). If Spitzer/IRAC 
shutter operations are allowed and successful, a carefully planned program should also be able to improve the absolute EBL
measurements at 3.6 and 4.5 $\mu$m over the coming years.

In Figure 2 we summarize EBL intensity measurements between optical and IR wavelengths using absolute photometry, 
model-dependent estimates based on the EBL fluctuations,
and the integrated galaxy light (IGL) from galaxy counts (lower limits). In general, the summed contribution of galaxies to
the EBL does not reproduce the EBL measured by absolute photometric
instruments.  For example, at $\lambda = 3.5 \, \mu$m the EBL measured
by DIRBE from absolute photometry is $13.0 \pm 4.8 \,$nW m$^{-2}$
sr$^{-1}$ (Levenson et al. 2007), while the deepest pencil beam surveys
with Spitzer at 3.6 $\mu$m give $6{-}9 \,$nW m$^{-2}$ sr$^{-1}$ (Fazio et al. 2004; Sullivan et al. 2007), with the
best determination of $9.0^{+1.7}_{-0.9} \,$nW m$^{-2}$ sr$^{-1}$ (Levenson \& Wright 2008).
At shorter wavelengths centered around 1 $\mu$m, and especially considering EBL measurements from IRTS (Matsumoto et al. 2005) 
this divergence is even more pronounced. 

Note that model-dependent fluctuations-based estimates of EBL
are in between IGL and absolute photometry measurements (at 1 $\mu$m, such estimates of EBL are from CIBER; Zemcov et al. 2014; 
Figure 2).
Fluctuation measurements and angular power spectra of source-subtracted 
optical and near-IR background intensity have been presented with Spitzer at 3.6 $\mu$m and above (Kashlinsky et al. 2005; Kashlinsky et al. 2012; Cooray et al. 2012), AKARI at 2.4, 3.2 and 4.1 $\mu$m (Matsumoto et al. 2011),
Hubble/NICMOS at 1.1 and 1.6 $\mu$m (Thompson et al. 2007), CIBER at 1.1 and 1.6 $\mu$m (Zemcov et al. 2014), and Hubble/ACS and WFC3 
in 5 bands from 0.6 and 1.6 $\mu$m (Mitchell-Wynne et al. 2015). These measurements generally reveal a picture in which source counts, mainly
galaxies, dominate the fluctuations with some evidence for additional components such as intra-halo light (IHL; Cooray et al. 2012; Zemcov
et al. 2014), associated with diffuse stars in extended dark matter halos due to galaxy mergers and tidal stripping,
 and a signal from galaxies present during reionization at $z > 8$ (Mitchell-Wynne et al. 2015). Spitzer fluctuations
have been also cross-correlated with far-infrared maps from Herschel/SPIRE (Thacker et al. 2015) and X-ray from Chandra (Cappelluti et al.
2013).  The former provides allows a quantification of the total dust content as a function of redshift while the latter has been used
to argue for the presence of primordial direct collapse black holes (DCHBs; Yue et al. 2013). Over the coming
decade significant improvements in the study of near-IR and optical fluctuations will come from planned
cosmological missions such as Euclid and WFIRST. The small explorer SPHEREx (Dor\'e et al. 2014), recently selected
by NASA for a Phase A study, has the ability to conduct 3D intensity mapping of spectral lines such as H$\alpha$ at $z \sim 2$
and Ly-$\alpha$ at $ z > 6$ during reionization over large areas on the sky.

A leading possibility for the large difference between absolute photometry EBL and IGL
is likely an unsubtracted foreground component, such as an underestimate of the Zodiacal light signal (Mattila 2006). 
If this difference, however, is
real it would have significant implications given that the nature of the emission source must be diffuse and not point-like similar
to galaxies. Fortunately, there is also a third technique to measure the EBL.
Given the limitation of direct measurements possibly due to foregrounds, currently the best estimates of optical/near-IR EBL
come from this indirect technique that makes use of the absorbed TeV/GeV spectra to constrain the optical and infrared background
due to pair production (shaded region in Figure 2 from Abramowski et al. 2013). The modeling requires intrinsic spectrum for each blazar,
but since this is not observed or available, EBL is inferred through statistical techniques
that make use of a large sample of blazars over a wide range of cosmological distances. The existing measurements come from High Energy 
Stereoscopic System (HESS) array in Namibia (Aharonian et al. 2006; Abramowski et al. 2013)
and Fermi/LAT (Ackermann 2012a; Gong \& Cooray 2013). The discrepancy between absolute photometry and IGL-implied intensity is
less severe when comparing galaxy counts to the EBL inferred from absorbed TeV spectra. The measurements are such that
deep galaxy counts have effectively resolved all of the optical and near-IR photons to individual galaxies. The overall
uncertainties, however, are still that the measurements leave the possibility for small signals such as IHL an reionization
consistent with fluctuation measurements.

Given the large uncertainties, including systematics, with TeV measurements, and especially absolute photometry measurements,
it is crucial that we improve on the optical and near-IR EBL intensity level. It is also clear that
simply repeating an absolute photometry experiment like DIRBE or IRTS at 1 AU will not improve the current EBL spectrum at 
wavelengths less than 5 $\mu$m due to limitations coming from the foreground model. 
Improvements in EBL measurements will only come with a parallel improvement in our understanding of the IPD distribution
in the Solar system, if measurements are limited to 1 AU, or from observations that are conducted outside of the
zodiacal light dust cloud.  In Figure 3 we show the expected zodiacal light intensity as a function of the radial distance from the Sun,
based on the {\it in-situ} dust measurements from Pioneer 10; the dust density is dropping more rapidly than
the $r^{-1}$ profile expected from the Poynting-Robertson effect (Hanner et al. 1974). At distances
of Jupiter these dust density measurements suggest a decrease in the zodiacal light intensity of roughly two orders
of magnitude from the intensity level near Earth orbit. 
One possibility is the out-of-Zodi EBL measurements, such as the proposed piggy-back ZEBRA instrument
on a planetary spacecraft to Jupiter or Saturn distances, or to travel outside of the
ecliptic plane. Given the relatively small cost of the instrument necessary for the required observations
we encourage attempts to measure EBL as a by-product of planetary spacecraft that explore the outer Solar system (Cooray et al. 2010).

\begin{figure}[!t]
\begin{center}
\includegraphics[width=0.9\textwidth]{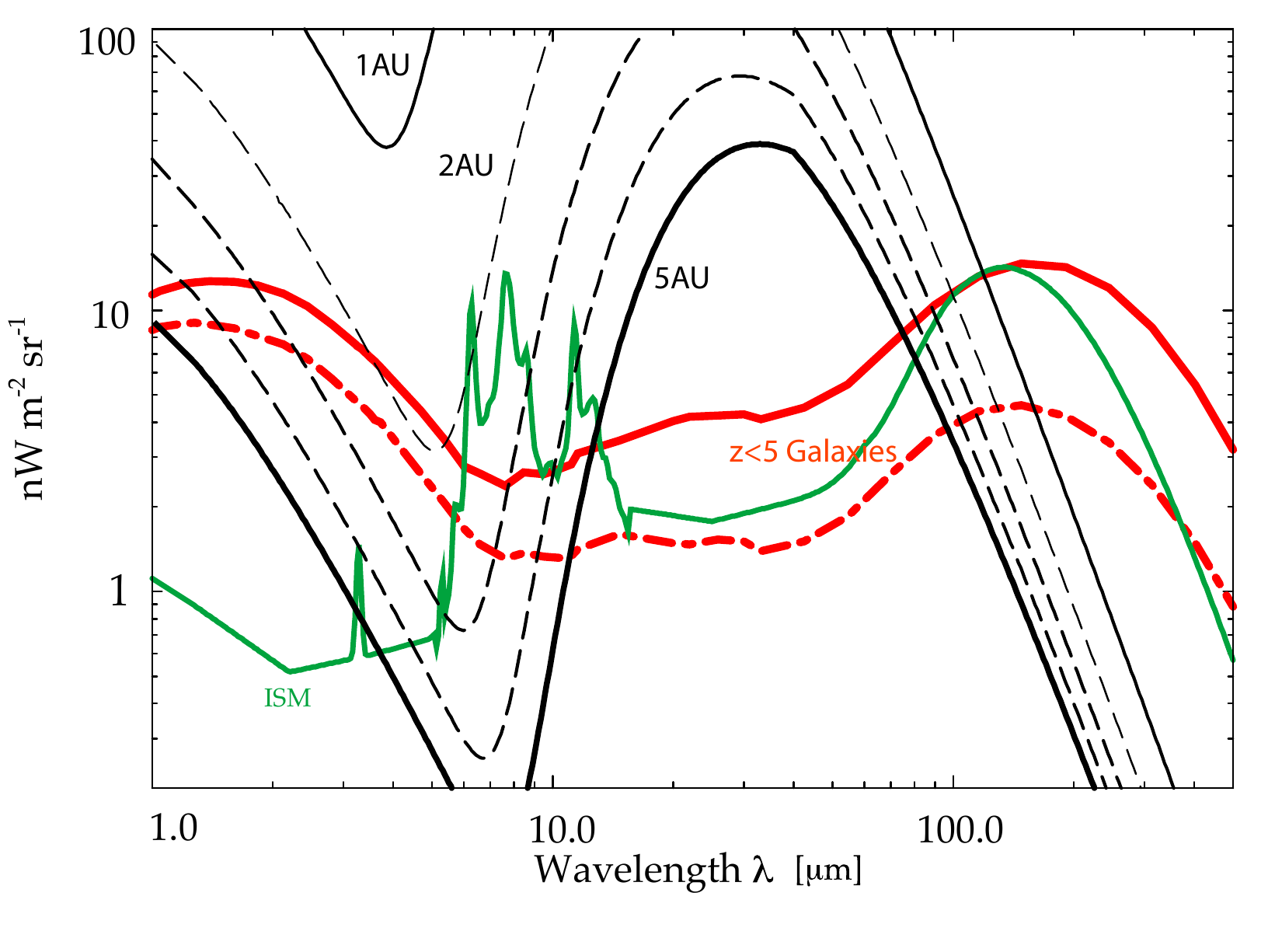}
\end{center}
\caption{
Zodiacal light intensity as a function of the radial distance from the Sun, based on the Pioneer 10 dust density estimates (Hanner et al.
2010) normalized to a radial profile from the Sun of $r^{-1.5}$. For reference we also show the optical to near-IR spectrum of the Galactic interstellar medium and two predictions from semi-analytical models on the total IGL from $z < 5$ galaxy populations (from Primack et al. 2008). The Zodical light intensity estimates as a function of the radial distance
is based out of the calculations for ZEBRA concept instrument for the outer Solar system in Cooray et al. (2009).
}
\end{figure}

\subsection{Far-Infrared}

Absolute photometry measurements between 10 $\mu$m and 1000 $\mu$m, corresponding to the far-infrared EBL peak at around 100 $\mu$m (Figure 2),
are mainly from DIRBE (Hauser et al. 1998) and FIRAS (Fixsen et al. 1998; Lagache et al. 2000). In general these measurements have
an overall uncertainty at the level of 30\%, all due to uncertain corrections associated with the foreground sky model.
This background is generally referred to as the cosmic infrared background (CIB) in some literature, especially in the
context of CMB experiments that also overlap in frequency ranges as the sub-millimeter wavelengths.

With the absolute photometry measurements establishing a cosmologically important energy density in the universe
at long infrared wavelengths, subsequent observations have focused on resolving this background to point sources with
large aperture telescopes. While the fraction resolved was low with Spitzer/MIPS at 70 and 160 $\mu$m, significant improvements
in our ability to detect and study distant galaxies came over the last five years with Herschel/PACS and SPIRE between 70 to 500 $\mu$m.
In particular, Herschel/SPIRE resolved 15\% (250 $\mu$m) to  5\% (500 $\mu$m) 
of the background directly to sources (Oliver et al. 2010). Using statistics such as
$P(D)$, probability of deflections, in deep SPIRE images, Glenn et al. (2010) resolved 
60\% (250 $\mu$m)  to 43\%  (500 $\mu$m) of the background to source counts, especially sources below the
individual point source detection level in maps. Methods involving stacking analysis resolve more of the background with recent
analysis suggesting a resolved fraction greater than 90\% (Viero et al. 2015). The sources that make up the far-infrared/submillimeter 
background are dusty, star-forming galaxies predominantly at high redshifts ($z>1$). They are likely the dominant contribution to the
cosmic star-formation rate density during the peak epoch of galaxy formation at $z \sim 2$-3. We refer the reader to the comprehensive
review by Casey et al. (2014) for properties of these galaxies.

Since COBE/DIRBE and FIRAS CIB intensity measurements the experimental 
focus has been on measurements related to the spatial anisotropies and the angular
power spectrum of CIB intensity (Haiman \& Knox 2000; Knox et al. 2001; Amblard \& Cooray 2007).  The power spectrum at 60 and 100 $\mu$m with IRAS
allowed studies related to the spatial distribution properties of Galactic dust and an estimate of
the total intensity of extragalactic sources through clustering and Poisson noise at small angular scales
(e.g., Miville-Desch\^nes et al. 2002). At 90 and 170 $\mu$m fluctuation measurements have also been attempted with ISO (Lagache \& Puget 2000;
Matsuhara et al. 2000). Significant improvements in our ability to remove Galactic emission and detect extragalactic fluctuations have
come from more recent experiments including Spitzer/MIPS at 160 $\mu$m (Lagache et al. 2007),
AKARI at 90 $\mu$ (Matsuura et al. 2011), and BLAST at 250, 350 and 500 $\mu$m (Viero et al. 2009).
Currently the best measurements of CIB power spectrum at a few degree to ten arcsecond angular scales are from Herschel/SPIRE (Amblard et al. 2010;
Viero et al. 2013; Thacker et al. 2013), while best measurements at larger angular scales and spanning the whole sky are from Planck (Ade et al. 2011; Ade et al. 2013a). 
While there was  a mismatch between Herschel/SPIRE  and Planck CIB power spectra at overlapping angular scales with first measurements
 this difference has mostly gone away with the latest flux calibration of Planck/HFI data. In combination, Planck and Herschel allow
studies of the CIB angular power spectrum from large linear scales with Planck to non-linear 1-halo term (Cooray \& Sheth 2002) with Herschel maps.
The measurements are useful to describe the spatial distribution of faint galaxies that make up the CIB and the relation between
far-IR luminosity to dark matter halo mass (Shang et al. 2011; De Bernardis \& Cooray 2012; Xia et al. 2012; Viero et al. 2013; Ade et al. 2013a).

Moving beyond the anisotropy power spectra, CIB fluctuations have also been used for cross-correlation studies. For example, far-IR galaxies
are correlated with unresolved near-IR background detected with Spitzer and the cross-correlation of near and far-IR backgrounds improve models
related to the dust distribution within dark matter halos (Thacker et al. 2015). Sources that make up CIB are mostly at $z> 1$. The foreground dark matter
potentials that are responsible for lensing of the CMB is mostly at $z \sim 1$-2 and CIB provides one of the best tracers of the projected dark matter related to CMB lensing
(Song et al. 2002). This cross-correlation of CMB lensing with CIB maps has been studied with Planck (Ade et al. 2013b) and
for detections of CMB lensing signal in the B-modes of CMB polarization (Hanson et al. 2013; Ade et al. 2014).

Additional future applications involving the far-IR/submm background include lensing of CIB fluctuations, that is CIB at $z \sim 3$ is expected to be 
gravitationally lensed by structures at $z < 1$, the search for a diffuse CIB components including intra-halo dust in dark matter halos that extend
beyond the dusty disks of star-forming galaxies, and a detailed statistical comparison of  dust in emission seen with CIB vs. dust seen in absorption through 
extinction studies (M\'enard et al. 2007). While the recent focus has been on CIB fluctuations and its applications,
the absolute CIB intensity still remains uncertain at the level of 30\% and must be improved down to sub-percent level
if detailed comparisons are to be made on the sources responsible for CIB vs. diffuse emission sources at far-IR wavelengths.
The proposed Primordial  Inflation Explorer (PIXIE; Kogut et al. 2014) has the capability to achieve such a measurement.
Its' sensitivity should also be adequate for a detection of the CIB dipole allowing a comparison of the CIB and CMB dipoles.

\subsection{Microwaves}

As shown in Figure~1, the cosmic microwave background (CMB), peaking at mm wavelengths between sub-mm and radio,
is the dominant background intensity across all wavelengths in the electromagnetic spectrum. Its total integrated
intensity of 960 nW m$^{-2}$ sr$^{-1}$ is roughly a factor of 30 higher than the integrated intensity of the infrared
background, which remains the next highest energetic component of the universe.
Since its' accidental discovery roughly 50 years ago, the high background intensity or photon energy density
has facilitated its wide applications in cosmology, especially with  spatial anisotropies and polarization
using a large number of ground and space-based experiments, including COBE, WMAP and Planck. For recent reviews on CMB theory
and experimental data summaries we refer the reader to Durrer (2015) and Komatsu et al. (2014).

While most of the experimental work in CMB concentrate on anisotropies and polarization, the
 best measurement of CMB spectrum, and correspondingly the best measurement of any EBL component from gamma-ray to radio, 
comes from COBE/FIRAS. It is described by a Planck function with a blackbody temperature
of $2.7260 \pm 0.0013$ (Fixsen 2009), with spectral departure from blackbody currently limited by the data to be at the level in $\delta I_\nu/I_\nu < 10^{-5}$ (Mather et al. 1994;
Fixsen et al. 1996). Distortions to the spectrum are expected at the micro and nano-Kelvin level (for a general review see Chluba 2014).
Detection and detailed study of these distortions, generated both during the early universe and at late times,
remain a primary scientific goal for a next generation CMB experiment, such as PIXIE (Kogut et al. 2014),  with sensitivity at least a factor of 30-100 better than FIRAS.

A well-known cosmological test related to the CMB temperature anisotropy power spectrum involves the location of the first acoustic peak in the multi-polar space (Kamionkowski et al. 1994). 
The CMB power spectrum from experiments like Planck now reveal the multiple acoustic peak structure  in the anisotropy power spectrum
from multipole moments 2 to $\sim$ 2500 and across at least eight peaks. Along with constraints on cosmological parameters (Ade et al. 2015a), these
observations now provide evidence for an initial spectrum of scale-invariant adiabatic density perturbations as expected under models involving
inflation. It has been argued for a while that the smoking-gun signature of inflation would be the detection of 
stochastic background of gravitational waves associated with it.  These gravitational-waves produce a distinct signature in the polarization of
CMB in the form of a contribution to the curl, or magnetic-like, component of the 
polarization (Kamionkowski et al. 1997). While polarization from density, or scalar, perturbations dominate,
due to the fact they have no handedness, there is no contribution to curl mode polarization from density perturbations.
Thus the current generation, and one focus of next generation measurements, involves CMB polarization and especially detailed characterization
of the B-modes of polarization.  

In transit to us, CMB photons also encounter the large scale structure that defines the local universe; thus, several 
aspects of photon properties, such as the frequency or the direction of propagation, are affected.
In the reionization epoch, variations are also 
imprinted when photons are  scattered via electrons, moving with respect to the CMB.
Though these secondary effects are in some cases insignificant compared to primary fluctuations, they
leave certain imprints in the anisotropy structure and induce higher order correlations. 
A well-studied example of such a secondary effect with current generation CMB experiments is lensing of the CMB (Lewis \& Challinor 2006),
with a significant detection of the lensing effect in Planck (Ade et al. 2015b).
The lensing of CMB is useful for cosmological applications involving structure formation and signatures leftover by a massive neutrinos.
A Stage IV CMB experiment will be able to reach the neutrino mass threshold expected given the neutrino oscillation experiments (Abazajian et al. 2015).
AS discussed with respect to Gamma-ray background, CMB lensing traces the large-scale structure that is also visible at other wavelengths.
Therefore, improvements in our understanding of the nature of dark matter and faint sources at each of the backgrounds will likely come from
cross-correlation studies. These are new topics in cosmology that will likely be improved over the coming years.
  
\subsection{Radio}

The cosmic radio background (CRB) has been measured at multiple frequencies and in recent years with the balloon-borne ARCADE-2 experiment from 3 to 90 GHz (Kogut et al. 2011).
When compared to CMB, Galactic synchrotron background, and extragalactic point sources, ARCADE 2 measured an excess radio background.
For example, at 3 GHz ARCADE measured the equivalent antenna temperature to be 65 mK, once corrected for CMB and Galactic emission. At the same frequency the known radio galaxy counts contribute
about 30 mK. If undetected radio sources account for the excess seen in ARCADE 2 they will need to form an extra peak in the Euclidean-normalized number
counts of radio sources at flux densities around 1-100 nJy, but an explanation involving point sources is ruled by various deep radio observations 
and other arguments (e.g., Condon et al. 2012; Holder 2014; Vernstrom et al. 2014). The excess has also motivated alternative suggestions, such as
decaying WIMP dark matter (e.g., Fornengo et al. 2011). A reanalysis of the Galactic synchrotron emission using multiple components instead of the single slab model for the Galactic plane
synchrotron emission used by the ARCADE team, however, suggests that there is likely no excess over the background produced by known sources (Subrahmanyan \& Cowsik 2013).
Future attempts to improve the radio background will thus likely also involve improvements to understanding and modeling of the Galactic radio foreground.

Current and next-generation experiments will likely focus more on the long wavelength radio background at frequencies around 100 MHz. These experiments
are driven by the need to characterize the background intensity spectrum to study the global 
signature associated with 21-cm spin-flip transition of HI from the epoch of reionization.  The global signal involves a strong absorption feature
around 60-100 mK, associated with adiabatic cooling of gas, followed by a weak emission during the epoch of reionization (Furlanetto 2006).
Detection of the expected absorption feature in the background intensity spectrum at frequencies around 60 MHz is challenging due to the large Galactic foreground at these low radio frequencies. Technology development studies are underway to pursue such a measurement from the Moon, including the lunar orbiter DARE (Dark Ages Radio Explorer; Jones et al. 2014).
There are also a host of experiments underway at frequencies above 100 MHz focused on the intensity fluctuations, especially the power spectrum of
21-cm background during reionization and for absolute measurements of the sky intensity (see review by Pritchard \&  Loeb 2012). 

\section{Summary}

This review covers the measurements related to the extragalactic background light (EBL) intensity
from gamma-rays to radio in the electromagnetic spectrum over 20 decades in the wavelength. 
The Cosmic Microwave Background (CMB) remains the best measured spectrum with an accuracy better than 1\%. 
The measurements related to the Cosmic Optical Background (COB), centered
at 1 $\mu$m, are impacted by the large Zodiacal light intensity  associated with interplanetary dust-scattered sunlight in the inner
Solar system. The best measurements of COB come from an indirect technique involving the absorption of
Gamma-ray photons emitted by bright blazars and other active sources in the universe.
The Cosmic Infrared Background (CIB) at wavelengths centered around 100 $\mu$m established
an energetically important intensity level comparable to the optical background. This eventually resulted
in the discovery of dusty, starbursting galaxies with large aperture telescopes and a deeper understanding of
their importance in galaxy formation and evolution. The soft X-ray/extreme UV extragalactic background at wavelengths of 10 to 100 nm 
remains mostly unexplored, but is unlikely to be achieved easily due to the absorption of the extragalactic photons
by the intervening neutral intergalactic medium and the interstellar medium of our Galaxy.
We also summarize our understanding of  spatial anisotropies of these backgrounds and the cosmological/astrophysical 
applications  with angular power spectra of intensity fluctuations across the sky.
We motivate a precise direct measurement of the COB between 0.1 to 5 $\mu$m  using a small aperture telescope
observing from either the outer Solar system or out of the ecliptic plane. Other future applications
include improving our understanding of the background at TeV energies, improving the MeV background over the previous
measurements with COMPTEL, radio background, and the spectral distortions to CMB and CIB.

\section*{Funding Statement}

My research is funded by NSF (CAREER AST-0645427, AST-1310310) and NASA.

\section*{Competing Interests}

I have no competing interests.

\section*{Acknowledgments}

I thank my collaborators in CIBER, SPHEREx, CANDLES, HerMES, H-ATLAS, ZEBRA, especially Jamie Bock, Bill Reach, Michael Zemcov,
Yan Gong, Ranga Chary, and many others for conversations related to
topics covered in this review. Nick Timmons and Ketron Mitchell-Wynne are acknowledged for help with figures and collection
of data in figures 1 and 2 from the literature to electronic format. The radial intensity of Zodiacal light in figure 3
comes from calculations performed by Bill Reach for the ZEBRA concept instrument in Cooray et al. (2009). Figure 2 is
an update to a similar figure from Dole et al. (2006) that discussed the measurements related to optical and infrared
extragalactic background light and the integrated light from galaxy/source counts.

\section*{Data Accessibility}

Electronic files listing the wavelength and intensity (in units of nW m$^{-2}$ sr$^{-1}$ as plotted in Figure 1) 
and the electronic versions of the figures are available from herschel.uci.edu.


\end{document}